\documentclass[11pt,a4paper]{article}
\usepackage[utf8]{inputenc}
\usepackage[margin=2.45cm]{geometry}
\usepackage{amsmath,amssymb}
\usepackage{graphicx}
\usepackage{booktabs}
\usepackage{natbib}
\usepackage{xcolor}
\usepackage{array}
\usepackage{caption}
\usepackage[hidelinks]{hyperref}

\usepackage{enumitem}

\title{\textbf{Before the Clinic: Transparent and Operable Design Principles for Healthcare AI}}

\author{
\href{https://orcid.org/0000-0001-7212-9573}{Alexander Bakumenko}\\
\textit{Clemson University}\\
Charleston, SC, USA\\
\texttt{abakume@clemson.edu}
\and
\href{https://orcid.org/0000-0002-2684-0548}{Aaron J Masino}\\
\textit{Clemson University}\\
Clemson, SC, USA\\
\texttt{amasino@clemson.edu}
\and
\href{https://orcid.org/0000-0003-3970-0613}{Janine Hoelscher}\\
\textit{Clemson University}\\
Clemson, SC, USA\\
\texttt{janineh@clemson.edu}
}

\date{}

\begin{document}

\maketitle

\begin{abstract}
The translation of artificial intelligence (AI) systems into clinical practice requires bridging fundamental gaps between explainable AI theory, clinician expectations, and governance requirements. While conceptual frameworks define what constitutes explainable AI (XAI) and qualitative studies identify clinician needs, little practical guidance exists for development teams to prepare AI systems prior to clinical evaluation. We propose two foundational design principles, Transparent Design and Operable Design, that operationalize pre-clinical technical requirements for healthcare AI. Transparent Design encompasses interpretability and understandability artifacts that enable case-level reasoning and system traceability. Operable Design encompasses calibration, uncertainty, and robustness to ensure reliable, predictable system behavior under real-world conditions. We ground these principles in established XAI frameworks, map them to documented clinician needs, and demonstrate their alignment with emerging governance requirements. This pre-clinical playbook provides actionable guidance for development teams, accelerates the path to clinical evaluation, and establishes a shared vocabulary bridging AI researchers, healthcare practitioners, and regulatory stakeholders. By explicitly scoping what can be built and verified before clinical deployment, we aim to reduce friction in clinical AI translation while remaining cautious about what constitutes validated, deployed explainability.
\end{abstract}

\section{Introduction}

The deployment of artificial intelligence (AI) in healthcare confronts a persistent translation gap. Despite sophisticated machine learning (ML) models demonstrating strong predictive performance in research settings, their adoption in clinical practice remains limited \cite{topol2019high, sendak2020real}. This gap persists not merely due to technical limitations but because of fundamental misalignments between what AI systems provide and what clinical environments require \cite{tonekaboni2019clinicians, elish2018stakes}.

Recent years have witnessed substantial progress in two complementary areas. First, conceptual frameworks for explainable AI (XAI) have matured, offering principled characterizations of interpretability, understandability, usability, and usefulness \cite{combi2022manifesto, lipton2018mythos}. Second, governance frameworks and reporting guidelines have emerged, specifying obligations for high-risk AI systems and standards for clinical trial reporting \cite{eu2019ethics, eu2024ai, vasey2022decide, liu2020consort}. Between these two bodies of work lies a critical gap: practical, pre-clinical guidance for research and 
development teams on what to build and verify prior to clinical evaluation. Clinicians should be engaged from project start as part of the development team; throughout this paper, "pre-clinical" refers to work completed before clinical evaluation or deployment, not before collaboration with clinicians during requirements elicitation (e.g., in Business and Data Understanding).

Combi et al. \cite{combi2022manifesto} propose that explainability emerges from the intersection of four characteristics: interpretability (intuiting causes of decisions), understandability (ascertaining how systems work), usability (ease of operation), and usefulness (practical worth). While conceptually comprehensive, this framework does not prescribe concrete engineering deliverables. Tonekaboni et al. \cite{tonekaboni2019clinicians} document what clinicians want from explainable systems (feature importance, uncertainty quantification, transparent design) through qualitative interviews, yet their work does not specify how development teams should technically prepare these artifacts prior to clinical evaluation and deployment. Governance frameworks such as the EU White Paper on AI \cite{eu2019ethics}, the EU AI Act \cite{eu2024ai} and NIST AI Risk Management Framework \cite{nist2023ai} define high-level characteristics of trustworthy AI (transparency, robustness, accountability) but remain intentionally sector-agnostic and principle-based rather than prescriptive.

The consequence is that development teams may proceed to clinical evaluation unprepared, lacking standardized approaches to pre-clinical XAI readiness. Reporting guidelines like DECIDE-AI \cite{vasey2022decide} and CONSORT-AI/SPIRIT-AI \cite{liu2020consort, cruz2021spirit} articulate what should be reported during early clinical studies and trials, but these frameworks assume that systems have already been developed with appropriate explanatory capabilities. Process models like CRISP-ML(Q) \cite{studer2021crisp} provide general machine learning workflows with quality assurance but lack healthcare-specific XAI artifacts such as modality attribution for multimodal models or calibrated fallback mechanisms for missing data.

\begin{figure}[t]
  \centering
  \includegraphics[width=\linewidth]{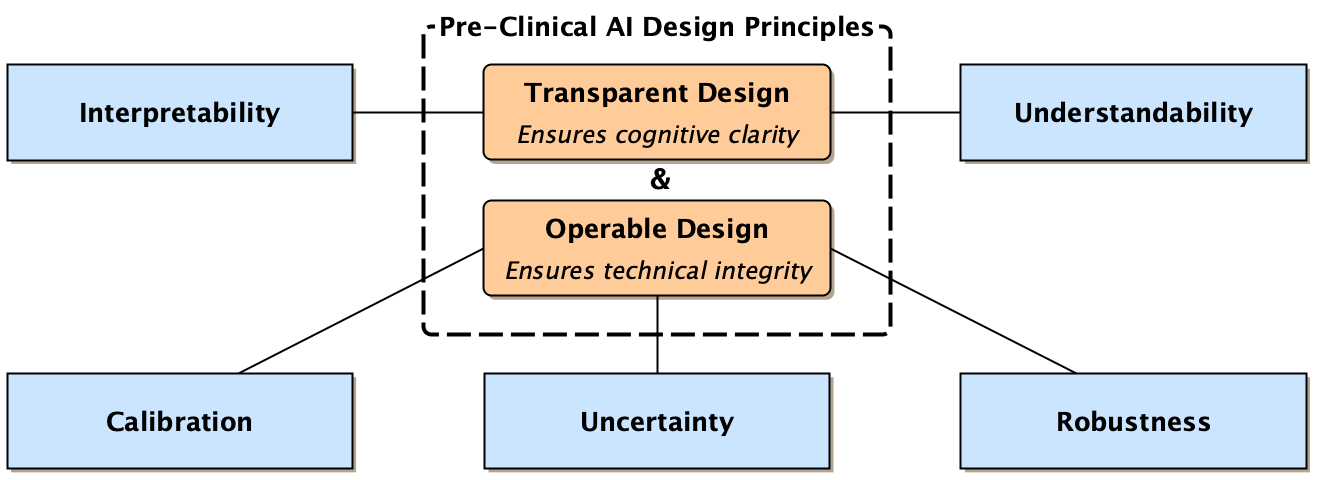}
  \caption{Pre-Clinical AI Design Principles. Transparent Design (Interpretability, Understandability) ensures cognitive clarity; Operable Design (Calibration, Uncertainty, Robustness) ensures technical integrity. The dashed rectangle marks the pre-clinical scope (work that can be built and verified prior to clinical evaluation). Connection lines denote component membership without causality or temporal order.}
  \label{fig:preclinical_principles}
\end{figure}

We address this gap by proposing two foundational principles, \textbf{Transparent Design} and \textbf{Operable Design}, as actionable pillars for pre-clinical healthcare AI development. These principles (framework pillars) operationalize technical requirements that can be built, tested, and documented before user studies or clinical trials, while explicitly acknowledging what cannot be claimed without such involvement. Our contribution is not to replace existing frameworks but to bridge them: connecting XAI theory to engineering practice, linking documented clinician needs to implementable artifacts, and preparing systems for governance compliance before deployment. We summarize the pre-clinical AI design principles and their components in Fig. \ref{fig:preclinical_principles}.

The development of this framework followed the logic of problematization as articulated by Alvesson et al. \cite{alvesson2013has}, which involves identifying, questioning, and reformulating assumptions in an existing domain of literature to generate alternative conceptual grounds. Consistent with Gregor’s classification of information systems theories \cite{gregor2006nature}, this work constitutes an "analysis"-type theory describing what can be built and verified before clinical evaluation rather than specifying causal or predictive relationships.

This paper is structured as follows. We first define Transparent Design (Section 2), encompassing interpretability and understandability artifacts that enable case-level reasoning and system traceability. We then characterize Operable Design (Section 3), addressing calibration, uncertainty, and robustness for reliable, predictable system behavior. Section 4 presents crosswalks mapping these principles to established frameworks (Combi's XAI components \cite{combi2022manifesto}, Tonekaboni's clinician needs \cite{tonekaboni2019clinicians}, EU governance requirements \cite{eu2019ethics}). We discuss limitations and appropriate handoffs (what should be passed on or transitioned) to clinical evaluation in Section 5, and conclude with implications for healthcare AI development in Section 6.

\section{Transparent Design: Interpretability and Understandability}

\subsection{Conceptual Foundation}

Transparent Design addresses the fundamental clinical need to understand why a system produces specific predictions and how it operates internally. Drawing from Combi et al. \cite{combi2022manifesto}, we distinguish two complementary but distinct characteristics that together constitute transparency artifacts:

\textbf{Interpretability} is "the degree to which a user can intuit the cause of a decision" and "the capability of predicting a system's result" \cite{combi2022manifesto}. In clinical contexts, interpretability enables practitioners to anticipate model behavior and identify case-level drivers of predictions without requiring deep knowledge of algorithmic internals. When an ICU clinician receives a mortality risk score, interpretability artifacts answer: \textit{Which patient characteristics drove this estimate?}

\textbf{Understandability} is "the degree to which a user can ascertain how the system works" and "being aware of how the system works" \cite{combi2022manifesto}. Unlike interpretability, understandability concerns the system's operational logic rather than case-specific reasoning. For instance, for multimodal models combining physiological measurements and clinical notes, understandability artifacts answer: \textit{How are data sources combined at the system level? What is the fusion mechanism?}

Combi et al. \cite{combi2022manifesto} emphasize that in knowledge-intensive medical tasks, distinguishing these concepts is crucial. In real-time decision support (e.g., responding to ICU alarms), interpretability may be critical as clinicians need immediate insight into which vital signs triggered alerts. In off-line analysis or system evaluation, understandability becomes essential. Understanding how the system derives results, including unexpected ones, enables validation against evolving medical knowledge.

Tonekaboni et al. \cite{tonekaboni2019clinicians} document that clinicians consistently request awareness of "the variables that have derived the decision of the model" and models that "reflect a similar analytic process to the established methodology of evidence-based medical decision making." One ICU clinician articulated: "would want to know the equation to know what the weights are." These needs map directly to our Transparent Design artifacts.

\subsection{Interpretability Artifacts: Case-Level Explanations}

Interpretability artifacts provide case-specific explanations of model predictions. 
We define three categories of artifacts that address different aspects of case-level reasoning:

\textbf{Feature Attribution.} For any given prediction, feature attribution methods identify which input features most influenced the outcome \cite{ribeiro2016why, lundberg2017unified, sundararajan2017axiomatic}. These methods range from model-agnostic approaches (e.g., LIME \cite{ribeiro2016why}, SHAP \cite{lundberg2017unified}) to model-specific techniques (e.g., Integrated Gradients \cite{sundararajan2017axiomatic} for neural networks, feature importance scores for tree-based  \cite{breiman2001random}). 

In healthcare contexts, feature attribution must be patient-specific \cite{tonekaboni2019clinicians}. Tonekaboni et al. found that clinicians expect to see both individual-level feature importance (what drove \textit{this patient's} risk score) and population-level patterns (what features generally matter). A junior clinician noted: "you have just a number, you can still use it but in your mind when you put all the variables that make you take a decision, the weight of that variable is going to be less than if you do understand exactly what that number means" \cite{tonekaboni2019clinicians}.

\textbf{Temporal Explanations.} For time-series clinical data, interpretability must capture how patient state changes over time influence predictions \cite{choi2016retain, tonekaboni2019clinicians}. In ICU settings, "clinicians are interested to see the change of state that has resulted in a certain prediction" \cite{tonekaboni2019clinicians}. This requires explanations that highlight critical time windows or trajectory shifts, for instance, identifying that a sustained decline in Glasgow Coma Scale scores \cite{teasdale1974assessment} over six hours preceded a high-risk prediction, rather than merely flagging instantaneous values.

\textbf{Modality Attribution (for Multimodal Systems).} When models integrate heterogeneous data sources (structured measurements, clinical notes, imaging, laboratory results), an additional layer of interpretability becomes necessary: \textit{which data modality dominated this specific decision?} For a model combining vital signs and clinical documentation, knowing which modality dominates a given high-risk prediction and to what extent (that is, quantifying the relative contributions) provides actionable insight. 
This per-case modality attribution enables hierarchical investigation aligned with clinicians' needs \cite{tonekaboni2019clinicians}: 
if notes dominated, which phrases were influential? If vitals dominated, which measurements and time periods?
Unlike feature attribution, which measures influence among individual variables within a modality, modality attribution quantifies how much each data source (e.g., clinical notes, images, vital signs) contributes to a specific prediction and precedes feature-level drill-down.

\subsection{Understandability Artifacts: System Traceability}

Understandability artifacts expose how the system operates globally, independent of specific cases. These artifacts enable validation, debugging, and governance.

\textbf{Transparent Fusion Mechanisms.} For ensemble or multimodal models, the method of combining component predictions should be inspectable \cite{tonekaboni2019clinicians}. Linear meta-learners or weighted voting schemes with explicit coefficients directly satisfy this requirement (stakeholders can examine how specialist model outputs are mathematically combined). Black-box fusion (e.g., deep neural network layers merging multimodal embeddings) sacrifices understandability even if individual components remain interpretable;
ensuring transparency in such architectures requires additional techniques and discussion beyond the scope of this paper.
Clinicians specifically request this transparency: "would want to know the equation to know what the weights are" \cite{tonekaboni2019clinicians}.

\textbf{Architecture Documentation and Lineage.} Complete system understandability requires documenting the model architecture, training procedures, data preprocessing pipelines, and versioning \cite{mitchell2019model}. This record must also describe the data used to develop the model, including patient demographics, relevant comorbidities, clinical characteristics, data sources and origins, and any relevant inclusion or exclusion criteria that define the development dataset. Model cards \cite{mitchell2019model} and datasheets \cite{gebru2021datasheets} provide standardized templates for this documentation. In healthcare, such documentation must additionally specify clinical context: intended patient populations, care settings, and known limitations.

\textbf{Global Feature Importance.} Complementing case-level attribution, global feature importance quantifies which features consistently influence predictions across the population \cite{breiman2001random, friedman2001greedy}. This addresses clinicians' need to understand model behavior in aggregate and validates alignment with clinical knowledge. If a sepsis prediction model assigns high importance to features unrelated to known pathophysiology, this misalignment can be detected through global importance analysis.

\subsection{Validation of Transparent Design Artifacts}

Transparent Design artifacts themselves require validation to ensure they reliably reflect model behavior. Two critical validation concerns have emerged from XAI evaluation research:

\textbf{Faithfulness.} Explanations should accurately represent how the model makes decisions, not merely provide plausible post-hoc rationalizations \cite{jacovi2020faithfulness}. Faithfulness can be assessed through deletion tests or perturbation studies. If an explanation identifies features as highly important, removing or masking those features should substantially alter predictions \cite{hooker2019benchmark}. Sanity checks \cite{adebayo2018sanity} ensure that explanation methods are truly linked to the trained model. Changing model parameters or training data should alter the explanations, rather than merely producing visually or textually salient patterns unrelated to the model's reasoning.

\textbf{Stability.} 
Complementing sanity checks, which verify sensitivity to meaningful model changes, stability focuses on consistency under small perturbations to inputs or parameters \cite{alvarez2018robustness}.
An explanation method that produces drastically different attributions for near-identical patients lacks reliability. Stability can be evaluated through bootstrap resampling or adversarial perturbations.

We emphasize that Transparent Design, while necessary, is not sufficient for deployment. These artifacts prepare systems for clinical engagement but do not replace user studies validating that explanations are \textit{usable} and \textit{useful} in practice \cite{combi2022manifesto, doshi2017towards}.

\section{Operable Design: Calibration, Uncertainty, and Robustness}
\subsection{Conceptual Foundation}

Operable Design addresses whether the system behaves predictably and reliably under real-world conditions. Importantly, operability concerns fall outside Combi et al.'s four XAI components (interpretability, understandability, usability, usefulness) but are nonetheless essential for trustworthy healthcare AI \cite{combi2022manifesto}. 

Combi et al. note that "reliability is a component of robustness that indicates the degree of trust placed in an ML model's prediction on a single example" \cite{combi2022manifesto}. The EU Ethics Guidelines for Trustworthy AI identify technical robustness and safety as a distinct requirement alongside transparency \cite{eu2019ethics, combi2022manifesto}. Thus, while Transparent Design addresses the explanatory surface clinicians see, Operable Design addresses the system's technical integrity, the foundation upon which explanations rest. In this work, we treat \emph{reliability} as the emergent outcome of three defined pre-clinical components of the operable design (calibration, uncertainty, and robustness), rather than as a standalone component.

Crucially, Tonekaboni et al. \cite{tonekaboni2019clinicians} found that clinicians perceive calibration and uncertainty quantification as part of explanation, even though these are technically robustness properties: "Presenting certainty score on model performance or predictions is perceived by clinicians as a sort of explanation that complements the output result." This perception underscores why pre-clinical preparation must address both transparency and operability. Clinicians do not separate them, and neither should development teams.

Unlike Transparent Design, requiring separate validation of its artifacts for faithfulness and stability, Operable Design components are intrinsically quantitative and include their own validation metrics (e.g., calibration error, coverage accuracy, robustness tests).

\subsection{Calibration: Aligning Predictions with Observed Frequencies}

Calibration ensures that predicted probabilities reflect true outcome frequencies \cite{degroot1983comparison}. 
A well-calibrated model’s predicted risks should correspond closely to the observed event rates across cases receiving that risk estimate.
Poor calibration undermines clinical utility: if predicted risks are systematically overconfident or underconfident, clinicians cannot appropriately set action thresholds \cite{guo2017calibration}. Multiple metrics quantify calibration quality:

\begin{itemize}[itemsep=3pt, parsep=3pt]
\item \textit{Expected Calibration Error (ECE)} measures the weighted average difference between predicted probabilities and observed frequencies across probability bins \cite{naeini2015calibration}.
\item \textit{Brier score} combines calibration and discrimination into a single proper scoring rule \cite{brier1950verification}.
\item \textit{Calibration plots} (reliability diagrams) visualize predicted probabilities versus observed frequencies, with perfect calibration forming a diagonal line \cite{degroot1983comparison}.
\item \textit{Calibration slope and intercept} from logistic recalibration quantify systematic over- or under-confidence \cite{cox1958regression, steyerberg2010assessing}.
\end{itemize}

When models exhibit miscalibration, post-hoc methods can improve reliability without retraining \cite{platt1999probabilistic, zadrozny2002transforming}. Platt scaling \cite{platt1999probabilistic}, temperature scaling \cite{guo2017calibration}, and isotonic regression \cite{zadrozny2002transforming} map predicted probabilities to calibrated estimates using held-out validation data. These methods are particularly valuable for neural networks, which often produce overconfident predictions \cite{guo2017calibration}.

Tonekaboni et al. \cite{tonekaboni2019clinicians} emphasize that "calibration of complex models" is "a significant technical challenge that needs to be addressed for clinical practice," and that clinicians overwhelmingly report that "clinical alignment in their judgment often determined their sustained use and trust in the model." This alignment requires not just high discrimination (AUROC) but accurate probability estimates.

\subsection{Uncertainty}

Beyond point estimates (single predictions), clinicians require awareness of prediction uncertainty \cite{tonekaboni2019clinicians, begoli2019need}. There are two main types of uncertainty:

\textbf{Aleatoric Uncertainty} results from inherent data noise or patient complexity. Some cases are inherently difficult to predict given available information \cite{kendall2017uncertainties}. Communicating aleatoric uncertainty helps clinicians recognize ambiguous cases requiring additional evaluation.

\textbf{Epistemic Uncertainty} reflects model limitations. The model may be uncertain because it has insufficient training data for this patient subgroup or because input features are outside the training distribution \cite{kendall2017uncertainties}. Epistemic uncertainty signals when models should abstain from making predictions rather than outputting unreliable estimates.

While aleatoric and epistemic uncertainty describe distinct sources of unpredictability, the system must handle both within a unified uncertainty-aware reporting policy. To operationalize these uncertainty estimates for practical decision-making, \emph{conformal prediction} (a framework producing prediction sets) \cite{vovk2005algorithmic, angelopoulos2021gentle} provides a formal basis for generating predictions with finite-sample coverage guarantees, optionally enabling principled abstention policies.
When uncertainty exceeds acceptable thresholds, the system can defer to human judgment rather than producing potentially harmful predictions. This uncertainty-informed reporting policy can be validated pre-clinically, for example through coverage–error or coverage–precision curves, while the specific operating point is selected collaboratively with clinicians during evaluation.

In practice, satisfying \emph{uncertainty} component requires measuring model uncertainty, identifying its source, defining how the system should respond, and validating that these behaviors are reliable before clinical use.

\subsection{Robustness: Predictable Behavior Under Distribution Shift}

Clinical AI systems must maintain performance despite inevitable deviations from training conditions \cite{finlayson2021clinicians, subbaswamy2019preventing}. Three robustness concerns stand out:

\textbf{Missing Data Robustness.} Healthcare data exhibits pervasive missingness with complex patterns: missing completely at random (MCAR), missing at random (MAR), or missing not at random (MNAR) \cite{little2019statistical}. Tonekaboni et al. note that "characterizing consistency under missingness... needs to be rigorously adopted and evaluated for clinical applications" \cite{tonekaboni2019clinicians}. Systems should be stress-tested across missingness patterns. When multiple data modalities are integrated, deterministic fallback mechanisms should be defined. For instance, in a multimodal system with two modalities, if clinical notes (first modality) are unavailable, the system should gracefully degrade to physiological measurements (second modality) alone, rather than failing catastrophically.

\textbf{Subgroup Performance.} Models may perform differentially across patient subgroups defined by demographics, disease subtypes, or care settings \cite{obermeyer2019dissecting}. Pre-clinical evaluation should stratify performance metrics by relevant subgroups, documenting disparities and characterizing populations where the model is or is not appropriate \cite{rajkomar2018ensuring}.
When clinical relevance of subgroup boundaries is uncertain, these should be identified or confirmed in consultation with domain experts during problem understanding.

\textbf{Temporal and Geographic Shift.} Clinical data distributions evolve due to changing patient populations, treatment protocols, or data collection practices \cite{finlayson2021clinicians}. While comprehensive drift adaptation requires post-deployment monitoring \cite{davis2006relationship}, pre-clinical assessment should evaluate model sensitivity to temporal splits (training on older data, testing on recent data) and geographic splits (training at one institution, testing at another) where feasible.

\subsection{Designing for Future Monitoring}

We do not include monitoring as a component of the operable design within our pre-clinical scope, as monitoring occurs post-deployment. However, pre-clinical preparation can enable future monitoring by defining basic infrastructure. This includes (i) recording model inputs, predictions, and true clinical outcomes for later review, (ii) tracking key performance indicators such as accuracy and calibration over time, and (iii) detecting when new data begin to differ from the training data \cite{davis2006relationship, quinonero2008dataset}. These monitoring foundations need to be informed by EU AI Act post-market surveillance obligations \cite{eu2024ai} and FDA Good Machine Learning Practice guidance on continuous learning \cite{fda2021gmlp}, ensuring regulatory alignment.

\section{Connecting Transparent and Operable Design Principles to Established Frameworks}

Transparent design and Operable Design do not exist in isolation. These two pillars bridge established theoretical frameworks, documented clinical needs, and emerging governance requirements. 

\begin{table*}[ht]
\centering
\caption{Mapping Transparent and Operable Design principles to established frameworks.
}
\label{tab:crosswalk}
\small
\begin{tabular}{>{\raggedright\arraybackslash}p{2.3cm}>{\raggedright\arraybackslash}p{4.0cm}>{\raggedright\arraybackslash}p{4.4cm}>{\raggedright\arraybackslash}p{3.6cm}}
\toprule
\textbf{Principle} & \textbf{Combi et al.’s XAI Component \cite{combi2022manifesto}} & \textbf{Tonekaboni et al.’s Clinician Need \cite{tonekaboni2019clinicians}} & \textbf{EU Trustworthy AI \cite{eu2019ethics}} \\
\midrule
\textbf{Transparent Design:} Feature attribution, modality attribution, transparent fusion & 
\textbf{Interpretability:} Enables intuiting causes of decisions and predicting system results. \textbf{Understandability:} Reveals how the system works, especially critical for off-line analysis. \cite{combi2022manifesto} & 
\textbf{Feature awareness:} "knowing the subset of features deriving the model outcome is crucial... to compare model decision to their clinical judgment." \textbf{System's Transparency:} "would want to know the equation to know what the weights are." \cite{tonekaboni2019clinicians} & 
\textbf{Transparency and Accountability:} Systems should provide clear information on their capabilities, limitations, and decision logic; documentation and record-keeping must allow traceability and human oversight \cite{eu2019ethics}. 
\\
\midrule
\textbf{Operable Design:} Calibration, uncertainty, missing-data robustness & 
\textbf{Reliability (component of robustness):} "indicates the degree of trust placed in an ML model's prediction on a single example." Reliability is not one of the four XAI components; it falls under technical robustness. \cite{combi2022manifesto} & 
\textbf{Uncertainty awareness:} "Presenting certainty score... is perceived by clinicians as a sort of explanation that complements the output result." 
\textbf{Calibration:} is "a significant technical challenge that needs to be addressed for clinical practice."\cite{tonekaboni2019clinicians}
& 
\textbf{Technical Robustness and Safety:} EU's second requirement for trustworthy AI. 
Systems must be accurate, resilient to errors, and behave reliably throughout their lifecycle \cite{eu2019ethics}.
\\
\bottomrule
\end{tabular}
\end{table*}

\clearpage
Table \ref{tab:crosswalk} presents crosswalk mapping our principles to three key references: Combi et al.'s XAI framework \cite{combi2022manifesto}, Tonekaboni et al.'s clinician needs \cite{tonekaboni2019clinicians}, and EU Trustworthy AI requirements \cite{eu2019ethics}. This mapping provides a shared vocabulary for interdisciplinary teams.

Transparent Design addresses two of Combi's four XAI components. Interpretability and understandability are necessary but not sufficient for Combi et al.'s full definition of explainability, which requires the intersection of interpretability, understandability, usability, and usefulness \cite{combi2022manifesto}. Pre-clinical work cannot claim usability or usefulness without user studies. These characteristics must be evaluated with clinicians in operational contexts \cite{combi2022manifesto, doshi2017towards, schoonderwoerd2021human}. Thus, we deliberately use "Transparent Design" rather than claiming "explainability."

Operable Design extends the robustness dimension beyond the XAI framework. Combi et al. position reliability as a component of robustness, distinct from their four XAI characteristics \cite{combi2022manifesto}. The EU White Paper similarly identifies "technical robustness and safety" as key requirements alongside transparency \cite{eu2019ethics}. Tonekaboni et al. document that clinicians perceive uncertainty and calibration as explanatory \cite{tonekaboni2019clinicians}. They do not conceptually separate XAI from reliability. This highlights why both pillars (transparent and operable design) must be addressed together: theoretical distinctions matter for precise communication, but clinical utility requires integrated preparation.

Both pillars align with EU trustworthy AI requirements. The EU White Paper links trustworthiness to transparency (clear information on system capabilities and limitations), robustness and accuracy (ensuring systems behave reliably), and appropriate human oversight \cite{eu2019ethics}. Transparent Design supports transparency and oversight obligations, while Operable Design supports robustness and accuracy expectations. Together, they are consistent with and support the EU’s "ecosystem of trust".

The conceptual mappings above show how our design principles align with existing frameworks and regulatory values. In practice, Transparent and Operable Design must also integrate into the machine learning lifecycle. CRISP-ML(Q) \cite{studer2021crisp} provides a six-phase machine learning process model: Business Understanding, Data Understanding, Data Preparation, Modeling, Evaluation, and Deployment. Transparent Design artifacts primarily emerge during Evaluation (explanations, faithfulness tests) but require planning during Business Understanding (which stakeholders need what transparency?) and Modeling (selecting architectures suited to explanation). Operable Design spans Evaluation (calibration assessment). Both principles benefit from explicit consideration throughout the lifecycle rather than applied post-hoc.

\section{Scope, Boundaries, and Handoffs}

\subsection{Framework Scope and Boundaries}

Transparent and Operable Design principles establish technical readiness but do not replace clinical evaluation. We acknowledge several important limitations of this scope.

Usability and usefulness require user studies. 
Combi et al. \cite{combi2022manifesto} emphasize that usability ("ease with which a user can learn to operate, prepare inputs for, and interpret outputs") and usefulness ("practical worth or applicability") are essential components of explainability. These characteristics cannot be claimed based solely on technical artifacts as they require evaluation with users in realistic workflows \cite{doshi2017towards, schoonderwoerd2021human}. Pre-clinical preparation provides candidates for useful explanations, but whether clinicians find them helpful in practice remains an empirical question.

Documenting subgroup performance disparities (Operable Design) is necessary but not sufficient to ensure fairness \cite{obermeyer2019dissecting, rajkomar2018ensuring}. Fairness requires normative judgments about acceptable tradeoffs, stakeholder engagement to define fairness criteria, and potentially algorithmic interventions beyond monitoring. Pre-clinical subgroup analysis reveals disparities but does not resolve them.

Clinical impact requires trials. Demonstrating that an AI system with transparent and operable design improves patient outcomes, clinician efficiency, or care quality requires rigorous evaluation \cite{vasey2022decide, liu2020consort}. Pre-clinical readiness can accelerate progression to such studies but does not substitute for them.

Applying XAI methods at different levels of the system involves inherent limitations. Common issues include the sensitivity of gradient-based attributions (e.g., saliency maps) to implementation details, which may not reliably reflect model reasoning \cite{adebayo2018sanity}; inconsistencies in attention-based explanations \cite{jain2019attention}; and the dependence of model-agnostic methods such as LIME on local approximation quality, which may fail in high-dimensional or non-smooth spaces \cite{slack2020fooling}. Pre-clinical preparation should acknowledge these limitations in documentation, apply faithfulness and stability tests where feasible.

\subsection{Appropriate Handoffs to Clinical Evaluation}

Our framework explicitly defines handoff points where pre-clinical work transitions to clinical engagement.

DECIDE-AI \cite{vasey2022decide} provides reporting guidelines for early-stage clinical evaluation of AI decision support systems. Once Transparent and Operable Design artifacts are prepared, DECIDE-AI guides assessment of usability, workflow integration, and preliminary safety signals. Development teams should use our framework to ensure systems are ready for DECIDE-AI evaluation, not as a replacement for it.

SPIRIT-AI and CONSORT-AI \cite{cruz2021spirit, liu2020consort} extend trial protocol and reporting standards to AI interventions. These guidelines cover study design, statistical analysis plans, and trial reporting for randomized evaluations. Pre-clinical documentation generated under our framework (e.g., interpretability and understandability artifacts, quantitative evaluation of system behavior, reports) directly supports SPIRIT-AI/CONSORT-AI protocol elements. It provides the technical basis for detailed intervention descriptions (how the AI system works and is intended to be used) and the conditions under which model performance is evaluated.

Human-Centered Design Studies \cite{barda2020qualitative} assess whether explanation interfaces genuinely support clinical decision-making. Pre-clinical explanation artifacts provide starting points for design iteration that precedes formal usability testing.

\section{Discussion and Implications}

\subsection{Significance of Pre-Clinical Principles}

The absence of standardized pre-clinical XAI guidance creates inefficiencies and risks. Development teams may invest heavily in explanation methods that fail to address clinician needs \cite{tonekaboni2019clinicians}, deploy poorly calibrated models that undermine trust \cite{guo2017calibration}, or reach clinical trials unprepared for governance scrutiny \cite{eu2024ai}. Conversely, teams may delay clinical engagement excessively, seeking unattainable perfection in explanation quality when iterative co-design with clinicians would be more productive \cite{yang2020re}. Transparent and Operable Design principles mitigate these risks by:

\begin{itemize}
\item Establishing shared vocabulary across AI researchers, clinicians, and regulators, reducing miscommunication about system capabilities.
\item Providing actionable targets for development teams: specific artifacts to build, tests to conduct, and documentation to maintain.
\item Accelerating governance readiness by aligning pre-clinical work with regulatory requirements (EU transparency, robustness) before deployment.
\item Enabling early risk identification through faithfulness tests, calibration assessment, and subgroup performance analysis, catching issues before clinical trials.
\item Defining clear handoffs to clinical evaluation phases, helping teams recognize when pre-clinical work is sufficient and user engagement should begin.
\end{itemize}

\subsection{Flexibility Across Implementation Choices}

Our principles intentionally avoid prescribing specific algorithms or explanation techniques. Transparent Design may employ interpretability methods such as SHAP values \cite{lundberg2017unified}, Integrated Gradients \cite{sundararajan2017axiomatic}, attention mechanisms \cite{choi2016retain}, or concept activation vectors \cite{kim2018interpretability}, among other suitable techniques, together with understandability practices such as transparent fusion documentation or model cards \cite{mitchell2019model}. The critical requirement is that some validated approach provides interpretability and understandability artifacts appropriate to the model architecture and clinical context.

Similarly, Operable Design does not mandate specific calibration methods, uncertainty quantification techniques, or robustness testing procedures. Teams should select approaches suited to their model types, computational constraints, and deployment environments. The principle requires that calibration and uncertainty be assessed and documented, and that robustness under missing data or distribution shift be evaluated with defined fallback strategies. The specific implementations may vary.

This flexibility is essential for healthcare AI diversity: risk prediction models, diagnostic classifiers, treatment recommendation systems, and image analysis applications have different explanation needs and technical constraints. A rigid checklist would either be too specific (excluding valid alternatives) or too general (providing little guidance). Transparent and Operable Design offer middle-ground principles. It is specific enough to be actionable, and flexible enough to accommodate diverse implementations.

\subsection{Limitations and Future Work}

Our framework synthesizes existing literature and provides conceptual organization. Several directions warrant further development. Empirical validation through longitudinal studies tracking multiple AI development projects would strengthen understanding of how these principles affect clinical translation outcomes and adoption patterns. Implementing both Transparent and Operable Design requires expertise in XAI methods, calibration techniques, and evaluation frameworks, and smaller development teams or resource-constrained settings may struggle to address all elements. Prioritization guidance (identifying which artifacts are most critical in which contexts), for example, could enhance practical applicability. Governance requirements, clinical expectations, and XAI techniques continue to evolve rapidly. The principles we propose reflect current understanding and are intended as a living framework, subject to refinement and extension as the field matures. Finally, we note our deliberate choice of the term "Operable", reflecting lexical awareness. It avoids the inconsistent use of related terms like \emph{reliability}, \emph{utility}, \emph{robustness} and \emph{usability} across domains, while "Transparent" remains aligned with established XAI terminology.

\section{Conclusion}

The translation of AI systems into clinical practice requires bridging theory, practice, and governance. Transparent and Operable Design principles provide actionable pre-clinical guidance for healthcare AI development teams. These principles define what can and should be built prior to clinical evaluation. By grounding them in established XAI frameworks, mapping them to documented clinician needs, and aligning them with emerging governance requirements, we aim to accelerate responsible healthcare AI development and deployment.

Transparent Design encompasses interpretability and understandability artifacts that enable case-level reasoning and system traceability. 
Operable Design addresses calibration, uncertainty, and robustness, ensuring predictable behavior under real-world conditions. 
Together, these two pillars prepare systems for clinical evaluation while recognizing the limits of what can be achieved before clinical testing. 

We call for adoption, critique, and refinement of these principles by the healthcare AI community. Researchers should empirically evaluate whether systems prepared under this framework lead to better clinical translation outcomes. The path from algorithm to clinical deployment remains challenging. Transparent and Operable Design principles offer a pragmatic roadmap for the critical first steps that development teams can take today, prior to clinical evaluation.

\bibliographystyle{plain}

\begin{thebibliography}{10}

\bibitem{adebayo2018sanity}
Julius Adebayo, Justin Gilmer, Michael Muelly, Ian Goodfellow, Moritz Hardt, and Been Kim.
\newblock Sanity checks for saliency maps.
\newblock {\em Advances in neural information processing systems}, 31, 2018.

\bibitem{alvarez2018robustness}
David Alvarez-Melis and Tommi~S Jaakkola.
\newblock On the robustness of interpretability methods.
\newblock {\em arXiv preprint arXiv:1806.08049}, 2018.

\bibitem{alvesson2013has}
Mats Alvesson and Jörgen Sandberg.
\newblock Has management studies lost its way? ideas for more imaginative and innovative research.
\newblock {\em Journal of management studies}, 50(1):128--152, 2013.

\bibitem{angelopoulos2021gentle}
Anastasios~N Angelopoulos and Stephen Bates.
\newblock A gentle introduction to conformal prediction and distribution-free uncertainty quantification.
\newblock {\em arXiv preprint arXiv:2107.07511}, 2021.

\bibitem{barda2020qualitative}
Alon~J Barda, Christopher~M Horvat, and Harry Hochheiser.
\newblock A qualitative research framework for the design of user-centered displays of explanations for machine learning model predictions in healthcare.
\newblock {\em BMC Medical Informatics and Decision Making}, 20(1):1--16, 2020.

\bibitem{begoli2019need}
Edmon Begoli, Tanmoy Bhattacharya, and Dimitri Kusnezov.
\newblock The need for uncertainty quantification in machine-assisted medical decision making.
\newblock {\em Nature Machine Intelligence}, 1(1):20--23, 2019.

\bibitem{breiman2001random}
Leo Breiman.
\newblock Random forests.
\newblock {\em Machine learning}, 45(1):5--32, 2001.

\bibitem{brier1950verification}
Glenn~W Brier.
\newblock Verification of forecasts expressed in terms of probability.
\newblock {\em Monthly Weather Review}, 78(1):1--3, 1950.

\bibitem{choi2016retain}
Edward Choi, Mohammad~Taha Bahadori, Jimeng Sun, Joshua Kulas, Andy Schuetz, and Walter Stewart.
\newblock Retain: An interpretable predictive model for healthcare using reverse time attention mechanism.
\newblock {\em Advances in neural information processing systems}, 29, 2016.

\bibitem{combi2022manifesto}
Carlo Combi, Beatrice Amico, Riccardo Bellazzi, Andreas Holzinger, Jason~H Moore, Marinka Zitnik, and John~H Holmes.
\newblock A manifesto on explainability for artificial intelligence in medicine.
\newblock {\em Artificial Intelligence in Medicine}, 133:102423, 2022.

\bibitem{cox1958regression}
David~R Cox.
\newblock The regression analysis of binary sequences.
\newblock {\em Journal of the Royal Statistical Society Series B: Statistical Methodology}, 20(2):215--232, 1958.

\bibitem{davis2006relationship}
Jesse Davis and Mark Goadrich.
\newblock The relationship between precision-recall and roc curves.
\newblock pages 233--240, 2006.

\bibitem{degroot1983comparison}
Morris~H DeGroot and Stephen~E Fienberg.
\newblock The comparison and evaluation of forecasters.
\newblock {\em Journal of the Royal Statistical Society: Series D (The Statistician)}, 32(1-2):12--22, 1983.

\bibitem{doshi2017towards}
Finale Doshi-Velez and Been Kim.
\newblock Towards a rigorous science of interpretable machine learning.
\newblock {\em arXiv preprint arXiv:1702.08608}, 2017.

\bibitem{elish2018stakes}
Madeleine~Clare Elish.
\newblock The stakes of uncertainty: developing and integrating machine learning in clinical care.
\newblock In {\em Ethnographic Praxis in Industry Conference Proceedings}, volume 2018, pages 364--380. Wiley Online Library, 2018.

\bibitem{eu2019ethics}
{European Commission}.
\newblock White paper on artificial intelligence: A european approach to excellence and trust, 2020.

\bibitem{eu2024ai}
{European Union}.
\newblock Regulation (eu) 2024/1689 of the european parliament and of the council on harmonised rules on artificial intelligence (artificial intelligence act).
\newblock Official Journal of the European Union, 2024.

\bibitem{finlayson2021clinicians}
Samuel~G Finlayson, Adarsh Subbaswamy, Karandeep Singh, John Bowers, Annabel Kupke, Jonathan Zittrain, Isaac~S Kohane, and Suchi Saria.
\newblock The clinician and dataset shift in artificial intelligence.
\newblock {\em New England Journal of Medicine}, 385(3):283--286, 2021.

\bibitem{friedman2001greedy}
Jerome~H Friedman.
\newblock Greedy function approximation: a gradient boosting machine.
\newblock {\em Annals of statistics}, pages 1189--1232, 2001.

\bibitem{gebru2021datasheets}
Timnit Gebru, Jamie Morgenstern, Briana Vecchione, Jennifer~Wortman Vaughan, Hanna Wallach, Hal~Daum{\'e} Iii, and Kate Crawford.
\newblock Datasheets for datasets.
\newblock {\em Communications of the ACM}, 64(12):86--92, 2021.

\bibitem{gregor2006nature}
Shirley Gregor.
\newblock The nature of theory in information systems.
\newblock {\em MIS quarterly}, pages 611--642, 2006.

\bibitem{guo2017calibration}
Chuan Guo, Geoff Pleiss, Yu~Sun, and Kilian~Q Weinberger.
\newblock On calibration of modern neural networks.
\newblock In {\em International conference on machine learning}, pages 1321--1330. PMLR, 2017.

\bibitem{hooker2019benchmark}
Sara Hooker, Dumitru Erhan, Pieter-Jan Kindermans, and Been Kim.
\newblock A benchmark for interpretability methods in deep neural networks.
\newblock {\em Advances in neural information processing systems}, 32, 2019.

\bibitem{jacovi2020faithfulness}
Alon Jacovi and Yoav Goldberg.
\newblock Towards faithfully interpretable nlp systems: How should we define and evaluate faithfulness?
\newblock {\em arXiv preprint arXiv:2004.03685}, 2020.

\bibitem{jain2019attention}
Sarthak Jain and Byron~C Wallace.
\newblock Attention is not explanation.
\newblock In {\em Proceedings of the 2019 Conference of the North American Chapter of the Association for Computational Linguistics: Human Language Technologies}, volume~1, pages 3543--3556, 2019.

\bibitem{kendall2017uncertainties}
Alex Kendall and Yarin Gal.
\newblock What uncertainties do we need in bayesian deep learning for computer vision?
\newblock {\em Advances in neural information processing systems}, 30, 2017.

\bibitem{kim2018interpretability}
Been Kim, Martin Wattenberg, Justin Gilmer, Carrie Cai, James Wexler, Fernanda Viegas, and Rory Sayres.
\newblock Interpretability beyond feature attribution: Quantitative testing with concept activation vectors (tcav).
\newblock In {\em International Conference on Machine Learning}, pages 2668--2677. PMLR, 2018.

\bibitem{lipton2018mythos}
Zachary~C Lipton.
\newblock The mythos of model interpretability: In machine learning, the concept of interpretability is both important and slippery.
\newblock {\em Queue}, 16(3):31--57, 2018.

\bibitem{little2019statistical}
Roderick~JA Little and Donald~B Rubin.
\newblock {\em Statistical analysis with missing data}, volume 793.
\newblock John Wiley \& Sons, 2019.

\bibitem{liu2020consort}
Xiaoxuan Liu, Samantha~Cruz Rivera, David Moher, Melanie~J Calvert, Alastair~K Denniston, Hutan Ashrafian, Andrew~L Beam, An-Wen Chan, Gary~S Collins, Ara DarziJonathan~J Deeks, et~al.
\newblock Reporting guidelines for clinical trial reports for interventions involving artificial intelligence: the consort-ai extension.
\newblock {\em The Lancet Digital Health}, 2(10):e537--e548, 2020.

\bibitem{lundberg2017unified}
Scott~M Lundberg and Su-In Lee.
\newblock A unified approach to interpreting model predictions.
\newblock {\em Advances in neural information processing systems}, 30, 2017.

\bibitem{mitchell2019model}
Margaret Mitchell, Simone Wu, Andrew Zaldivar, Parker Barnes, Lucy Vasserman, Ben Hutchinson, Elena Spitzer, Inioluwa~Deborah Raji, and Timnit Gebru.
\newblock Model cards for model reporting.
\newblock pages 220--229, 2019.

\bibitem{naeini2015calibration}
Mahdi~Pakdaman Naeini, Gregory Cooper, and Milos Hauskrecht.
\newblock Obtaining well calibrated probabilities using bayesian binning.
\newblock In {\em Proceedings of the AAAI conference on artificial intelligence}, volume~29, 2015.

\bibitem{nist2023ai}
{National Institute of Standards and Technology}.
\newblock Artificial intelligence risk management framework (ai rmf 1.0).
\newblock Technical report, U.S. Department of Commerce, 2023.

\bibitem{obermeyer2019dissecting}
Ziad Obermeyer, Brian Powers, Christine Vogeli, and Sendhil Mullainathan.
\newblock Dissecting racial bias in an algorithm used to manage the health of populations.
\newblock {\em Science}, 366(6464):447--453, 2019.

\bibitem{platt1999probabilistic}
John Platt et~al.
\newblock Probabilistic outputs for support vector machines and comparisons to regularized likelihood methods.
\newblock {\em Advances in large margin classifiers}, 10(3):61--74, 1999.

\bibitem{quinonero2008dataset}
Joaquin Quionero-Candela, Masashi Sugiyama, Anton Schwaighofer, and Neil~D Lawrence.
\newblock {\em Dataset shift in machine learning}.
\newblock The MIT Press, 2009.

\bibitem{rajkomar2018ensuring}
Alvin Rajkomar, Michaela Hardt, Michael~D Howell, Greg Corrado, and Marshall~H Chin.
\newblock Ensuring fairness in machine learning to advance health equity.
\newblock {\em Annals of Internal Medicine}, 169(12):866--872, 2018.

\bibitem{ribeiro2016why}
Marco~Tulio Ribeiro, Sameer Singh, and Carlos Guestrin.
\newblock " why should i trust you?" explaining the predictions of any classifier.
\newblock In {\em Proceedings of the 22nd ACM SIGKDD international conference on knowledge discovery and data mining}, pages 1135--1144, 2016.

\bibitem{cruz2021spirit}
Samantha~Cruz Rivera, Xiaoxuan Liu, An-Wen Chan, Alastair~K Denniston, Melanie~J Calvert, Hutan Ashrafian, Andrew~L Beam, Gary~S Collins, Ara Darzi, Jonathan~J Deeks, et~al.
\newblock Guidelines for clinical trial protocols for interventions involving artificial intelligence: the spirit-ai extension.
\newblock {\em The Lancet Digital Health}, 2(10):e549--e560, 2020.

\bibitem{schoonderwoerd2021human}
Tjeerd~AB Schoonderwoerd, Wietske Jorritsma, Mark~A Neerincx, and Karel Van Den~Bosch.
\newblock Human-centered xai: Developing design patterns for explanations of clinical decision support systems.
\newblock {\em International Journal of Human-Computer Studies}, 154:102684, 2021.

\bibitem{sendak2020real}
Mark Sendak, Madeleine~Clare Elish, Michael Gao, Joseph Futoma, William Ratliff, Marshall Nichols, Armando Bedoya, Suresh Balu, and Cara O'Brien.
\newblock " the human body is a black box" supporting clinical decision-making with deep learning.
\newblock In {\em Proceedings of the 2020 conference on fairness, accountability, and transparency}, pages 99--109, 2020.

\bibitem{slack2020fooling}
Dylan Slack, Sophie Hilgard, Emily Jia, Sameer Singh, and Himabindu Lakkaraju.
\newblock Fooling lime and shap: Adversarial attacks on post hoc explanation methods.
\newblock In {\em Proceedings of the AAAI/ACM Conference on AI, Ethics, and Society}, pages 180--186, 2020.

\bibitem{steyerberg2010assessing}
Ewout~W Steyerberg, Andrew~J Vickers, Nancy~R Cook, Thomas Gerds, Mithat Gonen, Nancy Obuchowski, Michael~J Pencina, and Michael~W Kattan.
\newblock Assessing the performance of prediction models: a framework for traditional and novel measures.
\newblock {\em Epidemiology}, 21(1):128--138, 2010.

\bibitem{studer2021crisp}
Stefan Studer, Thanh~Binh Bui, Christian Drescher, Alexander Hanuschkin, Ludwig Winkler, Steven Peters, and Klaus-Robert M{\"u}ller.
\newblock Towards crisp-ml (q): a machine learning process model with quality assurance methodology.
\newblock {\em Machine learning and knowledge extraction}, 3(2):392--413, 2021.

\bibitem{subbaswamy2019preventing}
Adarsh Subbaswamy and Suchi Saria.
\newblock From development to deployment: dataset shift, causality, and shift-stable models in health ai.
\newblock {\em Biostatistics}, 21(2):345--352, 2020.

\bibitem{sundararajan2017axiomatic}
Mukund Sundararajan, Ankur Taly, and Qiqi Yan.
\newblock Axiomatic attribution for deep networks.
\newblock In {\em International conference on machine learning}, pages 3319--3328. PMLR, 2017.

\bibitem{teasdale1974assessment}
Graham Teasdale and Bryan Jennett.
\newblock Assessment of coma and impaired consciousness: a practical scale.
\newblock {\em The lancet}, 304(7872):81--84, 1974.

\bibitem{tonekaboni2019clinicians}
Sana Tonekaboni, Shalmali Joshi, Melissa~D McCradden, and Anna Goldenberg.
\newblock What clinicians want: contextualizing explainable machine learning for clinical end use.
\newblock In {\em Machine learning for healthcare conference}, pages 359--380. PMLR, 2019.

\bibitem{topol2019high}
Eric~J Topol.
\newblock High-performance medicine: the convergence of human and artificial intelligence.
\newblock {\em Nature medicine}, 25(1):44--56, 2019.

\bibitem{fda2021gmlp}
{U.S. Food and Drug Administration}.
\newblock Good machine learning practice for medical device development: Guiding principles.
\newblock Technical report, U.S. Department of Health and Human Services, 2021.

\bibitem{vasey2022decide}
Baptiste Vasey, Myura Nagendran, Bruce Campbell, David~A Clifton, Gary~S Collins, Spiros Denaxas, Alastair~K Denniston, Livia Faes, Bart Geerts, Mudathir Ibrahim, et~al.
\newblock Reporting guideline for the early stage clinical evaluation of decision support systems driven by artificial intelligence: Decide-ai.
\newblock {\em bmj}, 377, 2022.

\bibitem{vovk2005algorithmic}
Vladimir Vovk, Alexander Gammerman, and Glenn Shafer.
\newblock {\em Algorithmic learning in a random world}.
\newblock Springer, 2005.

\bibitem{yang2020re}
Qian Yang, Aaron Steinfeld, Carolyn Rosé, and John Zimmerman.
\newblock Re-examining whether, why, and how human-ai interaction is uniquely difficult to design.
\newblock In {\em Proceedings of the 2020 CHI Conference on Human Factors in Computing Systems}, pages 1--13, 2020.

\bibitem{zadrozny2002transforming}
Bianca Zadrozny and Charles Elkan.
\newblock Transforming classifier scores into accurate multiclass probability estimates.
\newblock In {\em Proceedings of the eighth ACM SIGKDD international conference on Knowledge discovery and data mining}, pages 694--699, 2002.

\end{thebibliography}

\end{document}